\documentclass[a4paper,12pt,twoside]{article} 
\usepackage{a4wide}
\usepackage{graphicx}
\usepackage{rotating}

\def\ltap{\ \raisebox{-.4ex}{\rlap{$\sim$}} \raisebox{.4ex}{$<$}\ }
\def\gtap{\ \raisebox{-.4ex}{\rlap{$\sim$}} \raisebox{.4ex}{$>$}\ }

\newcommand{\deltaatm}{\mbox{$\Delta  m^2_{\mathrm{atm}} \ $}}
\newcommand{\deltasol}{\mbox{$ \Delta  m^2_{\odot} \ $}}

\newcommand{\eV}{\mbox{$ \  \mathrm{eV} \ $}}

\begin{document}

{\flushright
SISSA 38/2004/EP

hep-ph/0406106

}
\renewcommand{\thefootnote}{\fnsymbol{footnote}}
\begin{center}
\noindent{\Large \tt \bf 
On the Atmospheric Neutrino Oscillations, $\theta_{13}$ and
Neutrino Mass Hierarchy\footnote{Based on the invited talk
by S. T. Petcov at the NOON2004 International Workshop, February 11 -
15, 2004, Tokyo, Japan (to be published in the Proceedings of the
Workshop).}
}\vspace{4mm}

\noindent{\large
S.~T.~Petcov$^1$\footnote{Also at: Institute of Nuclear Research and
Nuclear Energy, Bulgarian Academy of Sciences, 1784 Sofia, Bulgaria.}
and Sergio Palomares-Ruiz$^{2,3}$ 
}\vspace{2mm}

\noindent{\small
$^1$ Scuola Internazionale Superiore di Studi Avanzati 
and Istituto Nazionale di Fisica Nucleare, I-34014 Trieste, Italy}
\\
$^2$ Department of Physics and Astronomy, UCLA,
  Los Angeles, CA 90095, USA
\\
$^3$ Department of Physics and Astronomy, Vanderbilt University,
  Nashville, TN 37235, USA

\end{center}
\vspace{4mm}
\vspace{6mm}
\renewcommand{\thefootnote}{\arabic{footnote}}
\setcounter{footnote}{0}

\begin{abstract}
We give predictions for the 
up-down asymmetry in the
Nadir angle dependence of the ratio $N_{\mu}/N_e$
of the rates of the $\mu-$like and $e-$like
multi-GeV events measured in water-\v{C}erenkov 
detectors (Super-Kamiokande, etc.)
in the case of 3-neutrino oscillations
of the atmospheric $\nu_e$ ($\bar{\nu}_e$)
and $\nu_\mu$ ($\bar{\nu}_\mu$), driven
by one neutrino mass squared difference,
$|\deltaatm| \equiv |\Delta m^2_{31}| \sim
(2.0 - 3.0)\times 10^{-3}~{\rm eV^2} 
\gg \Delta m^2_{21} \equiv \deltasol$.
This ratio is particularly sensitive
to the Earth matter effects in the 
atmospheric neutrino oscillations,
and thus to the values
of $\sin^2\theta_{13}$ and
$\sin^2\theta_{23}$, $\theta_{13}$ and
$\theta_{23}$ being the neutrino mixing angle
limited by the CHOOZ and Palo Verde experiments
and that responsible for the dominant 
atmospheric $\nu_\mu \rightarrow \nu_{\tau}$ 
($\bar{\nu}_\mu \rightarrow \bar{\nu}_{\tau}$)
oscillations, respectively. It is also sensitive  
to the type of neutrino mass 
spectrum which can be with normal 
($\deltaatm> 0$) 
or with inverted  
($\deltaatm < 0$) hierarchy.
\end{abstract}
\vspace{20pt}

\newpage
\section{Introduction}
 There has been a remarkable  progress in the studies of neutrino
oscillations in the last several years.
The experiments with solar, 
atmospheric and reactor neutrinos~\cite{sol,SKsolar,SNO1,SKatm,KamLAND} 
have provided  
compelling evidences for the 
existence of neutrino oscillations 
driven by nonzero neutrino masses and neutrino mixing.
Evidences for oscillations of neutrinos were
obtained also in the first long baseline
accelerator neutrino experiment K2K~\cite{K2K}.

   The latest addition to this magnificent effort
is the evidence 
presented at this Workshop by the Super-Kamiokande (SK) 
collaboration for an ``oscillation dip'' 
in the $L/E-$dependence,
of the (essentially multi-GeV) 
$\mu-$like atmospheric neutrino events~\cite{SKdip04}, 
$L$ and $E$ being the  
distance traveled by neutrinos and 
the neutrino energy.
This beautiful result represents the first 
ever observation of a direct effect 
of the oscillatory dependence 
on $L/E$ of the probability 
of neutrino oscillations in vacuum.

  An improved analysis of 
SK atmospheric neutrino data, performed recently 
by the SK collaboration, gave~\cite{SKatmo03} 
at 90\% C.L. 
\begin{equation} 
1.3 \,\times\, 10^{-3}\,\mbox{eV}^2\,\leq\,
|\deltaatm|\, \leq \,3.1\, \times\,10^{-3}\,\mbox{eV}^2~,
~~~~~~~0.90 \leq \sin^22\theta_{23} \leq 1.0~,
\label{atmo03a}
\end{equation}
%
\noindent with best fit values $|\deltaatm| \equiv \Delta m^2_{31} =
2.0\times 10^{-3}$ 
eV$^2$ and $\sin^2 2\theta_{23} = 1.0$ (see also
ref.~\cite{Fogliatm0308055}).   
Earlier analysis   
of the SK atmospheric neutrino 
data produced somewhat larger values of
$|\deltaatm|$: the  best fit value 
found, e.g., by SK collaboration~\cite{SKatmo03}
reads $|\deltaatm| = 2.5 \times 10^{-3} \ \eV$.
Finally, the values of
$|\deltaatm|$ and  $\sin^22\theta_{23}$,
deduced from the SK analysis of the 
$L/E$ dependence of the observed
$\mu-$like atmospheric neutrino 
events~\cite{SKdip04}, are 
compatible with the values
obtained in the other analysis.

   As is well-known, the 
atmospheric neutrino and K2K data
do not allow one to determine the signs
of $\deltaatm$ and of 
$\cos2\theta_{23}$ when
$\sin^22\theta_{23} \neq 1.0$.
The two possibilities,
$\deltaatm > 0$ and $\deltaatm < 0$,
correspond to two different
types of neutrino mass spectrum:
with normal hierarchy (NH),
$m_1 < m_2 < m_3$, and 
with inverted hierarchy (IH),
$m_3 < m_1 < m_2$.  The 
ambiguity in the sign  of
$\cos2\theta_{23}$ 
implies that when, e.g., 
$\sin^22\theta_{23} = 0.92$,
two values of  $\sin^2\theta_{23}$ are possible,
$\sin^2\theta_{23} \cong 0.64~{\rm or}~ 0.36$.

  A very important parameter in the phenomenology of 3-neutrino
mixing and oscillations is the 
 angle $\theta_{13}$, limited by the data
from the CHOOZ and Palo Verde
experiments. The precise limit on 
$\theta_{13}$ is $\deltaatm-$ dependent 
(see, e.g, ref.~\cite{BNPChooz}).
Using the 99.73\%  allowed range of 
$\deltaatm = (1.1 - 3.2)\times 10^{-3}~{\rm eV^2}$
from ref.~\cite{Fogliatm0308055}, 
one gets from  a combined 3-neutrino
oscillation analysis of the solar neutrino,
CHOOZ and KamLAND data~\cite{SNO3BCGPR}:
\begin{equation}
\sin^2 \theta_{13} < 0.047~(0.074),~~~~~~~~~ 90\%~(99.73\%)~{\rm C.L.}
\end{equation}
%
The global analysis of the
solar, atmospheric and reactor neutrino data 
performed in ref.~\cite{ConchaNOON04}
gives $\sin^2 \theta_{13} < 0.054$ at 99.73\% C.L.

   Getting more precise 
information about the value 
of the mixing angle $\theta_{13}$,
determining the sign of $\deltaatm$, or
the type of the neutrino mass spectrum 
(with normal or inverted hierarchy), 
and measuring the value of 
$\sin^2\theta_{23}$ with a 
higher precision
is of fundamental importance
for the progress in the 
studies of neutrino mixing. 

 In ref.~\cite{JBSP203} we 
have derived predictions for the Nadir angle ($\theta_n$)
dependence of the ratio $N_{\mu}/N_e$
of the rates of the $\mu-$like and $e-$like
multi-GeV events measured in water-\v{C}erenkov detectors
in the case of 3-neutrino oscillations
of the atmospheric $\nu_e$ ($\bar{\nu}_e$)
and $\nu_\mu$ ($\bar{\nu}_\mu$), driven
by one neutrino mass squared difference,
$|\deltaatm| \sim
(2.0 - 3.0)\times 10^{-3}~{\rm eV^2} 
\gg \deltasol$.
This ratio was shown to be
particularly sensitive
to the Earth matter effects in the 
sub-dominant  $\nu_{\mu} \leftrightarrow \nu_e$
($\bar{\nu}_{\mu} \leftrightarrow \bar{\nu}_e$)
oscillations of the atmospheric 
$\nu_{\mu}$ ($\bar{\nu}_{\mu}$) and
$\nu_e$ ($\bar{\nu}_e$)~\cite{SP3198,SPNu98},
and thus 
i) to the value
of $\sin^2\theta_{13}$ which drives
the subdominant oscillations,
ii) to the value of
$\sin^2\theta_{23}$ which
determines the maximal possible value of the
corresponding subdominant transition probabilities, 
and iii) to the type of neutrino mass spectrum,
i.e., the sign of $\deltaatm$.
It was shown in ref.~\cite{JBSP203}, in particular,
that for $\sin^2\theta_{13} \gtap 0.01$,
$\sin^2\theta_{23}\gtap 0.5$ and 
at $\cos\theta_n \gtap 0.4$,
the Earth matter effects 
modify substantially
the $\theta_n-$ dependence of 
the ratio $N_{\mu}/N_e$ and in a way 
which cannot be reproduced with   
$\sin^2\theta_{13} = 0$
and a different value of $\sin^2\theta_{23}$.
For normal hierarchy, the effects 
of interest can be as large as 
$\sim 25\%$ for $\cos\theta_n \sim (0.5 - 0.8)$,
can reach $\sim 35\%$ in the Earth core bin
$\cos\theta_n \sim (0.84 - 1.0)$,
and might be observable~\cite{JBSP203}.
They were shown to be typically by
$\sim 10\%$ smaller in the inverted hierarchy case. 
This permitted to conclude that 
an observation of the Earth matter effects 
in the Nadir angle distribution of the 
ratio $N_{\mu}/N_e$
would clearly indicate that
$\sin^2\theta_{13} \gtap 0.01$ and 
$\sin^2\theta_{23} \gtap 0.50$.

 In the present article we give predictions 
for the up-down (U-D) asymmetry (see also ref.~\cite{AkhDig}) 
in the Nadir angle 
dependence of the ratio $N_{\mu}/N_e$
of the rates of the 
$\mu-$like and $e-$like
multi-GeV events measured in water-\v{C}erenkov 
detectors (Super-Kamiokande, etc.).
As in ref.~\cite{JBSP203}, we consider
the case of 3-neutrino oscillations
of the atmospheric $\nu_e$ ($\bar{\nu}_e$)
and $\nu_\mu$ ($\bar{\nu}_\mu$), driven
by one neutrino mass squared difference,
$|\deltaatm| \sim
(2.0 - 3.0)\times 10^{-3}~{\rm eV^2} 
\gg \deltasol$. The indicated U-D 
asymmetry is expected to have 
substantially smaller 
systematic uncertainty
than (the Nadir angle dependence of) the ratio
$N_{\mu}/N_e$ itself.

\section{Effects of Subdominant 3-$\nu$ Oscillations} 

 The subdominant $\nu_{\mu} \rightarrow \nu_{e}$
($\bar{\nu}_{\mu} \rightarrow \bar{\nu}_{e}$)
and $\nu_{e} \rightarrow \nu_{\mu (\tau)}$
($\bar{\nu}_{e} \rightarrow \bar{\nu}_{\mu (\tau)}$)
oscillations of the multi-GeV
atmospheric neutrinos of interest should exist 
and their effects could be observable 
if three-flavor-neutrino 
mixing takes place in vacuum,
i.e., if $\sin^22\theta_{13} \neq 0$, 
and if $\sin^22\theta_{13}$ is
sufficiently large~\cite{SP3198,SPNu98,matter,core}
(see also ref.~\cite{3nuKP88}).  
These transitions are 
driven by $\deltaatm$.
The probabilities of these transitions contain
$\sin^2\theta_{23}$ as factor which determines
their maximal value.
For $\deltaatm> 0$, the
 $\nu_{\mu} \rightarrow \nu_{e}$ 
($\bar{\nu}_{\mu} \rightarrow \bar{\nu}_{e}$)
and $\nu_{e} \rightarrow \nu_{\mu (\tau)}$
($\bar{\nu}_{e} \rightarrow \bar{\nu}_{\mu (\tau)}$)
transitions of the multi-GeV 
atmospheric 
neutrinos (antineutrinos)
are amplified (suppressed) 
by the Earth matter effects; if
$\deltaatm < 0$, the transitions of 
$\nu_{\mu}$, $\nu_{e}$
are suppressed and those of 
$\bar{\nu}_{\mu}$, $\bar{\nu}_{e}$
are enhanced. Therefore for a given sign of 
$\deltaatm$, the Earth matter affects 
differently the transitions of neutrinos and antineutrinos.
Thus, the study of the subdominant atmospheric neutrino
oscillations can provide information, 
in particular, about the 
sign of $\deltaatm$ and
the magnitudes of 
$\sin^2\theta_{13}$ and $\sin^2\theta_{23}$. 
     
  Under the condition
$|\deltaatm| \gg \deltasol$,
the relevant three-neutrino 
$\nu_{\mu} \rightarrow \nu_{e}$ 
($\bar{\nu}_{\mu} \rightarrow \bar{\nu}_{e}$)
and $\nu_{e} \rightarrow \nu_{\mu (\tau)}$
($\bar{\nu}_{e} \rightarrow \bar{\nu}_{\mu (\tau)}$)
transition probabilities reduce effectively 
to a 2-neutrino transition probability~\cite{3nuSP88} with 
$\deltaatm$  and $\theta_{13}$ playing the role of the relevant
two-neutrino oscillation parameters.

The fluxes of atmospheric $\nu_{e,\mu}$ 
of energy $E$, which reach the detector after
crossing the Earth along a given trajectory  
specified by the value of $\theta_{n}$, 
$\Phi_{\nu_{e,\mu}}(E,\theta_{n})$, 
are given by the following expressions 
in the case of the three-neutrino oscillations  
under discussion~\cite{SPNu98}:
\begin{equation}
\Phi_{\nu_e}(E,\theta_{n}) \cong 
\Phi^{0}_{\nu_e}~\left (  1 + 
  [s^2_{23}r - 1]~P_{2\nu}\right ),
\label{Phie}
\end{equation}
$$ \Phi_{\nu_{\mu}}(E,\theta_{n}) \cong \Phi^{0}_{\nu_{\mu}} 
\left ( 1 +
 s^4_{23}~ [(s^2_{23}~r)^{-1} - 1]~P_{2\nu} \right. $$
\begin{equation} 
\left. - 2c^2_{23}s^2_{23}~\left [ 1 -
Re~( e^{-i\kappa}
A_{2\nu}(\nu_{\tau} \rightarrow \nu_{\tau})) \right ] \right ). 
\label{Phimu}
\end{equation}
\noindent Here $\Phi^{0}_{\nu_{e(\mu)}} = 
\Phi^{0}_{\nu_{e(\mu)}}(E,\theta_{n})$ is the
$\nu_{e(\mu)}$ flux in the absence of neutrino 
oscillations and
\vspace{-0.4cm}
\begin{equation}
r \equiv r(E,\theta_{n}) \equiv
\frac{\Phi^{0}_{\nu_{\mu}}(E,\theta_{z})} 
{\Phi^{0}_{\nu_{e}}(E,\theta_{z})}~,
\label{r}
\end{equation}
%
$P_{2\nu} \equiv 
P_{2\nu}(\deltaatm, \theta_{13};E,\theta_{n})$ 
is the probability of two-neutrino 
$\nu_{e} \rightarrow \nu'_{\tau}$
oscillations in the Earth,
where  $\nu'_{\tau} = s_{23}\nu_{\mu} + c_{23} \nu_{\tau}$,
and $\kappa$ and 
$A_{2\nu}(\nu_{\tau} \rightarrow \nu_{\tau})$
are known phase and 
2-neutrino transition probability
amplitude~\cite{3nuSP88,SP3198,SPNu98,JBSP203}. 

For the predicted
ratio $r(E,\theta_{n})$ of the atmospheric 
$\nu_{\mu}$ and $\nu_e$ fluxes 
for i) the Earth core crossing and ii) only 
mantle crossing neutrinos, 
having trajectories for which
$0.4 \ltap \cos\theta_{n}\leq 1.0$, one has~\cite{Honda}:
$r(E,\theta_{z}) \cong (2.0 - 2.5)$ for 
the neutrinos giving  
contribution to the sub-GeV
samples of Super-Kamiokande events, 
and $r(E,\theta_{n}) \cong (2.6 - 4.5)$ for those 
giving the main contribution to the multi-GeV samples.
If $s^2_{23} = 0.5$ and $r(E,\theta_{z}) \cong 2.0$,
we have $(s^2_{23}~r(E,\theta_{z}) - 1) \cong 0$,
$((s^2_{23}~r(E,\theta_{z}))^{-1} - 1) \cong 0$,
and the possible effects of the 
$\nu_{\mu} \rightarrow \nu_{e}$ 
and $\nu_{e} \rightarrow \nu_{\mu (\tau)}$ 
transitions on the $\nu_e$ and $\nu_{\mu}$ 
fluxes, and correspondingly in 
the sub-GeV $e-$like and $\mu-$like 
samples of events, 
would be strongly suppressed.
The effects of interest 
are much larger for the
multi-GeV neutrinos 
than for the sub-GeV neutrinos. They are also 
predicted to be larger for
the flux of (and event rate due to) multi-GeV  
$\nu_e$ than for the flux of (and event rate due to) 
multi-GeV $\nu_\mu$. 

  The same conclusions are valid for the 
effects of oscillations on the fluxes of, and 
event rates due to, atmospheric antineutrinos
$\bar{\nu}_e$ and $\bar{\nu}_{\mu}$.

   Equations~(\ref{Phie}) -~(\ref{Phimu}) 
and the similar equations for
antineutrinos imply that in the case under study 
the effects of the $\nu_{\mu} \rightarrow \nu_{e}$,
$\bar{\nu}_{\mu} \rightarrow \bar{\nu}_{e}$, 
and $\nu_{e} \rightarrow \nu_{\mu (\tau)}$,
$\bar{\nu}_{e} \rightarrow \bar{\nu}_{\mu (\tau)}$,
oscillations 
i) increase with the increase of $s^2_{23}$ and are maximal
for the largest allowed value of $s^2_{23}$,
ii) should be substantially larger in the multi-GeV 
samples of events than in the sub-GeV samples, and
iii) in the case of the multi-GeV samples, for 
$\deltaatm > 0$ ($\deltaatm < 0$)
they lead to an increase of the
rate of $e^-$ ($e^+$) 
events and to a decrease of the 
$\mu^{-}$ ($\mu^+$) event rate.
The last point follows from the fact that
the magnitude of the effects we are interested in 
depends also on the 2-neutrino oscillation probabilities,
$P_{2\nu}$ and $\bar{P}_{2\nu}$,
and that $P_{2\nu}$ or
$\bar{P}_{2\nu}$ (but not both probabilities) 
can be strongly enhanced  by the Earth matter effects. 

 A more detailed analysis shows (see, e.g.,
 refs.~\cite{JBSP203,PalPet}) 
that for $\deltaatm = (2 - 3)\times 10^{-3}~{\rm eV^2} > 0$,
the Earth matter effects can 
amplify $P_{2\nu}$
significantly when the neutrinos cross 
{\it only the Earth mantle} i) for 
$E \sim (6 - 11)$ GeV, 
and ii) only for sufficiently long 
neutrino paths in the mantle, i.e., for
$\cos\theta_n \gtap 0.4$.
The magnitude of the matter effects 
increases with increasing 
of $\sin^2\theta_{13}$. The energy 
$E_{res}$ and the path length of neutrinos in the mantle,
$L$, for which one can have $P_{2\nu} \cong 1$,
are determined by the conditions:
\begin{equation}
E_{res} \cong 6.6 \left(\frac{\deltaatm}{10^{-3}~{\rm eV^2}}\right)
\left(\frac{\rm {N_A cm^{-3}}}{\rm \bar{N}_e^{man}}\right)
  \cos2\theta_{13}~{\rm GeV}~,
\label{Eres}
\end{equation}
\vspace{-0.16cm}
\begin{equation}
 1.2~\tan2\theta_{13}~   
 \left(\frac{\rm{\bar{N}_e^{man}}}{\rm{N_A cm^{-3}}}\right)
  \left(\frac{L}{10^{4}~{\rm km}}\right) = 1 , 
\label{Xman}
\end{equation}
\noindent $\bar{N}_e^{man}$ and
$N_A$ being the mean electron number density along the 
neutrino trajectory in the mantle and the Avogadro
number.

  In the case of atmospheric 
neutrinos crossing the Earth core, new
resonance-like effects become apparent. 
For $\sin^2\theta_{13} < 0.05$ and
$\deltaatm > 0$,
 we can have $P_{2\nu} \cong 1$ 
{\it only due to the effect of maximal constructive 
interference between the amplitudes of the 
$\nu_{e} \rightarrow \nu'_{\tau}$
transitions in the Earth mantle and in the 
Earth core}~\cite{SP3198,106,107}.
The effect differs from the MSW one  and the 
enhancement happens in the case of interest at 
a value of the energy  between the resonance energies 
corresponding to the density in the mantle 
and that of the core~\cite{SP3198}.
The {\it mantle-core enhancement effect} 
is caused by the existence 
(for a given neutrino trajectory
through the Earth core) of 
{\it points of resonance-like 
total neutrino conversion}, 
$P_{2\nu} = 1$,
in the corresponding space 
of neutrino oscillation 
parameters~\cite{106,107}.
A rather complete set of values of 
$\deltaatm/E$ and $\sin^22\theta_{13}$
for which $P_{2\nu} = 1$
for the Earth core-crossing atmospheric 
$\nu_{\mu}$ and $\nu_{e}$
was found in ref.~\cite{107}. 
The location of these points 
determines the regions
where $P_{2\nu}$ 
is large, $P_{2\nu} \gtap 0.5$. 
For $\sin^22\theta_{13} < 0.10$,
there is one set of values of 
$\deltaatm/E$ and $\sin^2\theta_{13}$
for which $P_{2\nu} = 1$. This  ``solution''
occurs for, e.g., $\theta_n = 0;~13^0;23^0$,
and  $\deltaatm = 2.0~(3.0)\times 10^{-3}~{\rm eV^2}$,
at  $\sin^22\theta_{13} = 0.034;~0.039;~0.051$, 
and $E \cong (2.8 - 3.1)~{\rm GeV}$
($E \cong (4.2 - 4.7)~{\rm GeV}$),
see Table 2 in ref.~\cite{107}.

\section{Results}
\begin{figure}
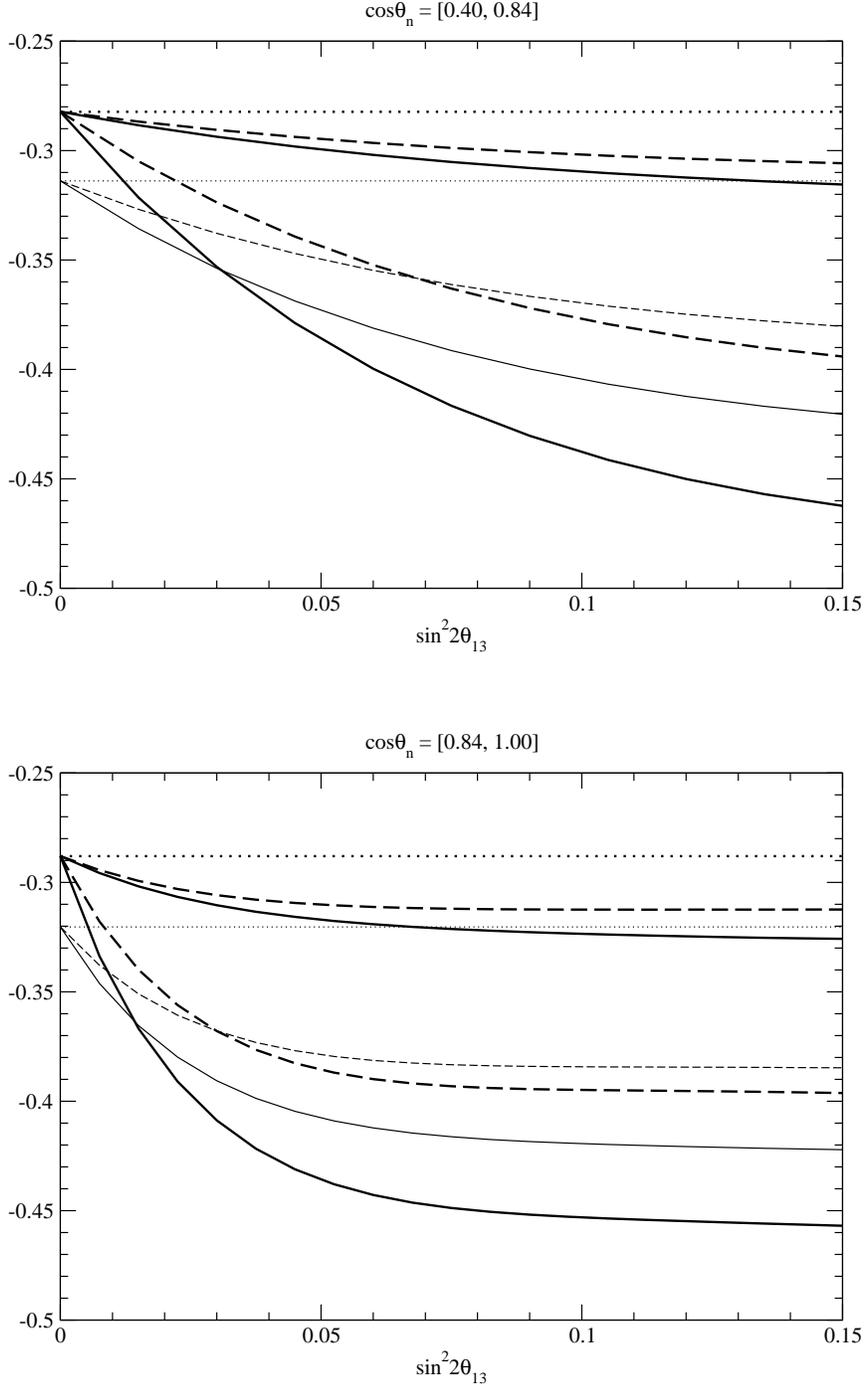

\begin{center}
\includegraphics[height=8.7cm]{UD2_04_084.eps}

\vspace{1.cm}

\includegraphics[height=8.7cm]{UD2_084_1.eps}
\end{center}
\caption{\label{fig:1} The up-down asymmetry in the Nadir angle 
dependence of the ratio $N_{\mu}/N_e$
of the rates of the $\mu-$like and $e-$like
multi-GeV events measured in water-\v{C}erenkov 
detectors, as a function of $\sin^22\theta_{13}$,
calculated for two intervals of values of
$\cos\theta_n$, [0.40,0.84] ({\it mantle bin} - upper panel)  
and  [0.84,1.0] ({\it core bin} - lower panel), 
for $|\deltaatm| = 2\times 10^{-3}~{\rm eV^2}$ and
i) $\deltaatm > 0$ - normal hierarchy (solid lines), 
ii) $\deltaatm < 0$ - inverted hierarchy (dashed lines), and 
iii) 2-neutrino vacuum oscillations (dotted lines),
and for $\sin^2\theta_{23} = $0.36 (upper thick lines),~
0.50 {\rm (thin lines)},~0.64 {\rm (lower thick lines)}. 
In the case of vacuum oscillations, 
there is no distinction between $\sin^2\theta_{23} = $0.36 and 0.64
(upper dotted line).
} 
\end{figure}
  We use the method of calculation of the 
up-down (U-D) asymmetry in the Nadir angle 
($\theta_n$) dependence of the ratio $N_{\mu}/N_e$
of the rates of the $\mu-$like and $e-$like
multi-GeV events, $A(U-D)$, 
measured in water-\v{C}erenkov 
detectors (Super-Kamiokande, 
Hyper-Kamiokande~\cite{HyperK}, etc.),
described in ref.~\cite{JBSP203}.
Our results are presented graphically   
in Fig. 1, where we show 
the asymmetry $A(U-D)$
as a function of $\sin^22\theta_{13}$,
calculated for two intervals of values of
$\cos\theta_n$, [0.40,084] (``mantle bin'') 
and [0.84,1.0] (``core bin''),
for $\deltaatm = \pm 2\times 10^{-3}~{\rm eV^2}$ 
and for $\sin^2\theta_{23} = 0.36;~0.50;~0.64$.
As Fig. 1 shows, for
$\sin^22\theta_{13}\ltap 0.06$, 
$A(U-D)$ in the {\it core bin} increases 
rapidly with $\sin^22\theta_{13}$,
and remains practically constant 
for $0.06 \ltap \sin^22\theta_{13} \leq 0.15$.
For  $\sin^2\theta_{23} = 0.64$
it reaches the values of (-0.45)
for $\deltaatm > 0$ (NH),
and (-0.39) if $\deltaatm < 0$ (IH)
at $\sin^22\theta_{13}= 0.06$, 
while the asymmetry in the case of 2$-\nu$
vacuum oscillations of the $\nu_{\mu}$
and $\bar{\nu}_{\mu}$ is 
considerably smaller in absolute value,
(-0.28). In the case of $\sin^2\theta_{23} = 0.50$,
the corresponding
asymmetry  values are
(-0.42), (-0.38) and (-0.32), respectively.
The asymmetry in the {\it mantle bin}
increases monotonically with the increase of
$\sin^22\theta_{13}$. At $\sin^22\theta_{13}= 0.06$ 
it is smaller than the asymmetry in the {\it core bin},
but at $\sin^22\theta_{13}= 0.15$
the asymmetries 
in the {\it mantle}
and the {\it core} bins 
practically coincide. 
For $\sin^2\theta_{23} \sim 0.36$,
the Earth matter effects 
in the subdominant neutrino oscillations
are suppressed and the U-D
asymmetries are essentially 
determined by their 2-neutrino
vacuum oscillation values.

  It is interesting to note that
using the SK atmospheric neutrino data~\cite{SKatmo03} 
one finds for the U-D asymmetry in the 
two mantle bins, $\cos\theta_n = [0.40,0.60],~
[0.60,0.84]$ and in the core bin, respectively:
$A_{m1}(U-D) = -0.29 \pm 0.13$, $A_{m2}(U-D) = -0.36\pm 0.14$,
$A_{c}(U-D) = -0.48\pm 0.16$.  
 
\section{Conclusions}

We have shown that the up-down asymmetry in the 
Nadir angle dependence of the ratio $N_{\mu}/N_e$
of the rates of $\mu-$like and $e-$like
multi-GeV events measured in water-\v{C}erenkov 
detectors (Super-Kamiokande, etc.)
is sensitive to the Earth matter effects in 
the subdominant oscillations  
of  the multi-GeV ($\sim (2 - 10)$ GeV) atmospheric neutrinos,
$\nu_{\mu} \rightarrow \nu_e$, $\nu_{e} \rightarrow \nu_{\mu}$,
$\bar{\nu}_{\mu} \rightarrow \bar{\nu}_e$ and 
$\bar{\nu}_{e} \rightarrow \bar{\nu}_{\mu}$.
The measurement with increased sensitivity of this
asymmetry can provide fundamental information on the values 
of $\sin^2\theta_{13}$ and $\sin^2\theta_{23}$, and 
on the sign of $\deltaatm$, i.e., on the neutrino mass hierarchy.

\section{Acknowledgments}

S.T.P. would like to thank the Organizers of the
NOON2004 International Workshop, and in particular
Y. Suzuki, for kind hospitality.
He is also indebted to T. Kajita and O. Yasuda 
for very useful discussions. This work was supported by  
the Italian INFN under the 
program ``Fisica Astroparticellare'' (S.T.P.)
and by the NASA grant NAG5-13399 (S.P.-R.).

\end{document}